\documentclass[doublecol]{epl2}

\title{Equilibration and symmetry breaking in vibrated granular systems }
\shorttitle{Title} 

\author{J. Javier Brey\inst{1} \and Nagi Khalil\inst{1}}
\shortauthor{F. Author \etal}

\institute{
  \inst{1} F\'{\i}sica Te\'{o}rica, Universidad de Sevilla, Apartado de Correos 1065,
E-41080 Sevilla, Spain}

\abstract{The steady states of two vibrated granular gases separated by an adiabatic piston are investigated.
The system exhibits a non-equilibrium phase transition with an spontaneous symmetry breaking. Even if the gases at both sides of the piston have the same number of particles and are mechanically identical, their steady volumes and temperatures can be rather different.
The transition can be explained by a simple kinetic theory model expressing mechanical equilibrium and  the energy balance occurring in the system. The model predictions are in good agreement with molecular dynamics simulation results. The macroscopic description of the steady states is discussed, as well as some physical implications of the symmetry breaking.
}

\begin{document}

\maketitle

The adiabatic piston is a prototypical construction to investigate the relaxation of a system and the relevance of fluctuations and successive dynamic time scales as the system evolves. It consists of a gas-filled container divided into two compartments by a freely moving adiabatic piston \cite{Ca63}. When the initial states of the gases in the two compartments are different, the relaxation can be rather complex even when the whole system is isolated and the final state is the equilibrium one \cite{Li99,KVyM00,GPyL03,Re02,GyL06}. Of course, the behavior of the system is much less understood when it can not relax to an equilibrium state due to some external constrains. In this case, even the answer to apparently simple questions is not known. For instance, is there any thermodynamic-like parameter characterizing the macroscopic equilibration between the two subsystems? If there is a final steady state, does it depend and in which way on the properties of the piston?

Granular media are a key type  of systems to address the above and many other issues related with fundamental concepts in far from equilibrium systems are granular media. Due to the inelasticity of  collisions, granular fluids are intrinsic non-equilibrium systems since no steady state is possible in an isolated system. On the other hand, the energy dissipation in collisions  offers the possibility of new energy balances and, consequently, steady states showing many peculiarities as compared with time-independent states of molecular, elastic systems are observed \cite{Ca90,Go03}.

In the case of granular gases, the dynamics is dominated by (inelastic) binary collisions. Therefore, it is not surprising that the techniques developed in kinetic theory and statistical mechanics for ordinary molecular fluids have been extended with success to them \cite{ByP04,BGyP09}. An advantage of granular gases over molecular fluids is that their smallest scale is macroscopic and, therefore, many of their characteristic features are directly observable in experiments.

The aim of the work being reported here is to study the steady state of two vibrated granular gases separated by an adiabatic piston. There are some previous studies of adiabatic pistons involving inelasticity effects. In some of them, the whole system is isolated, and the granular gases collide elastically with the piston. The system as a whole
is cooling, and the two granular gases are considered initially in the same macroscopic homogenous state. It has been shown that this system exhibits a phase transition with an spontaneous symmetry breaking \cite{BRyvB05,ByK10}. In other studies, infinite baths of elastic gases colliding inelastically with the piston have been considered. It is assumed that the baths are not affected by the piston, so that a steady state is reached after some transient
\cite{CMyP08,PTyB10}. Of course, all the above studies deal with highly idealized situations hard to approximate in real experiments. In particular, the homogeneous cooling state is of too short a duration to be observed. To overcome this limitation, in the present work, energy is continuously supplied to the gases in both compartments by vibrating the walls. Actually, the construction described below tries to mimic the experimental setup used in ref. \cite{LyD10}, although the results reported there can not be related to the theoretical analysis here, due to the rather different range of densities in both cases: very dilute here and very dense in \cite{LyD10}. In addition, no friction is considered in the present analysis. Very recently, the velocity and position fluctuations of a piston on top of a vibrated granular gas in presence of gravity have also been investigated, and a rather complex and not well understood behavior has been found \cite{ByR10}.

\begin{figure}
\onefigure[scale=0.3]{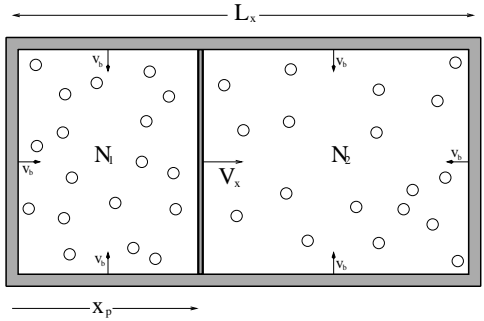}
\caption{Sketch of the system considered in this work. All the external walls are vibrating in a sawtooth way with velocity $v_{b}$.}
\label{fig1}
\end{figure}

We consider a cylindrical container of length $L_{x}$ divided into two compartments by a movable piston constrained to remain perpendicular to the axis of the system, taken as $x$ axis. The system is sketched in Fig. \ref{fig1}. The two compartments contain $N_{1}$ and $N_{2}$ smooth inelastic hard spheres ($d=3$) or disks ($d=2$), respectively. A subscript $i=1,2$ will be used to identify properties on each side of the piston. The particles within each subsystem are equal  and their mass and diameters will be denoted by $m_{i}$ and $\sigma_{i}$, $i=1,2$, respectively. Collisions between particles are characterized by velocity-independent coefficients of normal restitution $\alpha_{i}$, $0 <\alpha_{i} \leq 1$. On the other hand, collisions between particles and the piston, as well as collisions of particles with the walls of the container are taken as elastic. The piston moves without friction with the walls and it is modeled as having mass $M$ and negligible width. Energy is continuously supplied to the system through the external walls. All them vibrate with a sawtooth velocity profile and an amplitude much smaller than the mean free path of the particles in their vicinity. That means that to describe collisions, the walls can be considered as fixed and with a constant velocity $v_{b}$ perpendicular to them, and addressed towards the interior of the system \cite{McyLu98}.

The focus here will be on the steady state position $x_{P}$ eventually reached by the piston as a function of the parameters defining the system. This is equivalent to study the way in which the fixed total volume is divided into the two subsystems by the movable piston. In the following analysis, the rather drastic approximation that the granular gas inside  each of the compartments can be treated as homogeneous will be made. The expectation is that this simplification does not affect the qualitative behavior of the system, although it certainly will modify the accuracy of the quantitative predictions.  Moreover, attention will be restricted to those situations in which both gases are very dilute, so that their equations of state have the form $p_{i}=n_{i}T_{i}$, where $p_{i}$ is the pressure, $n_{i}$ the number density, and $T_{i}$ the granular temperature of the gas in compartment $i$. As usual, the latter is defined from the second moment of the velocity distribution with the Boltzmann constant formally set equal to unity. It is
\begin{equation}
\label{1}
n_{1} \equiv \frac{N_{1}}{S_{P} x_{P}}\, , \quad n_{2} \equiv \frac{N_{2}}{S_{P} \left( L_{x}-x_{P}\right)}\, ,
\end{equation}
where $S_{P}$ is the area ($d=3$) or length ($d=2$) of the piston and, therefore, also the section of the container. In the steady state, mechanical equilibrium of the piston requires that $p_{1}=p_{2}=p$, i.e.
\begin{equation}
\label{2} n_{1}T_{1}=n_{2}T_{2}.
\end{equation}
Although the stationarity of $x_{P}$ does not imply by itself the same for the temperatures or the pressure, only stationary steady sates of the whole system will be addressed in the following. Then energy balance for the gases in each of the two  compartments reads
\begin{equation}
\label{3}
\frac{d}{2} N_{i} \zeta_{i}T_{i}- Q_{Pi}S_{P}-Q_{Wi}S_{Wi}=0.
\end{equation}
The first term of the left hand side of this equation is the Haff law, describing the homogeneous cooling of the gas as a consequence of the inelasticity of collisions \cite{Ha83}. It involves the cooling rate $\zeta_{i}$ for which an accurate expression is \cite{GyS95,vNyE98}
\begin{equation}
\label{4}
\zeta_{i}= \frac{p \zeta^{*} (\alpha_{i})}{\eta_{0i}(T_{i})}\, ,
\end{equation}
where
\begin{equation}
\label{5}
\zeta^{*}(\alpha) = \frac{(2+d)(1-\alpha^{2})}{4d} \left[ 1+ \frac{3 c^{*}(\alpha)}{32} \right],
\end{equation}
\begin{equation}
\label{6}
c^{*}(\alpha)= \frac{32 (1- \alpha)(1-2 \alpha^{2})}{9+24d+(8d-41) \alpha+30 \alpha^{2} (1- \alpha)}\, ,
\end{equation}
and $\eta_{0i}(T)$ is the shear viscosity of a dilute molecular gas of hard spheres or disks,
\begin{equation}
\label{7}
\eta_{0i}(T) = \frac{d+2}{8}\, \Gamma \left( \frac{d}{2} \right) \pi^{-\frac{d-1}{2}} \left(m_{i} T \right)^{1/2} \sigma_{i}^{-(d-1)}\, .
\end{equation}
The second term in Eq.\ (\ref{3}) is the power going into compartment $i$ through the piston due to its velocity fluctuations. This is the way in which the gases at both sides of the piston interchange energy. To describe it, it is convenient to introduce a temperature parameter $T_{P}$ for the piston defined from the second moment of its velocity fluctuations, namely
\begin{equation}
\label{8}
M \langle V_{x}^{2} \rangle  \equiv T_{P},
\end{equation}
with the angular brackets denoting statistical average. For $m_{i}/M \ll 1$, $i=1,2$, and assuming that $T_{P}/T_{i}$ is of the order of unity, it is \cite{ByK10,ByR09a}
\begin{equation}
\label{9}
Q_{Pi} \approx -2 \left( \frac{2m_{i}}{\pi} \right)^{1/2} \frac{T_{i}-T_{P}}{M} n_{i}T_{i}^{1/2}.
\end{equation}
Finally, the last term in Eq.\ (\ref{3}) is the power injected into the subsystem $i$  by the vibrating walls in contact with it. For the specific kind of vibration considered here \cite{ByR09a},
\begin{equation}
\label{10}
Q_{Wi}=p v_{b}.
\end{equation}
Moreover, $S_{Wi}$ is the total section (area or length) of the vibrating walls in contact with the gas in compartment $i$.  This quantity is a function of $x_{P}$. The steady energy balance for the piston reads
\begin{equation}
\label{11}
Q_{P1}+Q_{P2}=0.
\end{equation}
Let us introduce dimensionless temperature parameters by
\begin{equation}
\label{12}
\theta_{i} \equiv \frac{T_{i}}{m v_{b}^{2}}\, , \quad
\theta_{P} \equiv \frac{T_{P}}{m v_{b}^{2}}\, ,
\end{equation}
where
\begin{equation}
\label{13}
m \equiv \frac{m_{1} m_{2}}{m_{1}+ m_{2}}\, ,
\end{equation}
and also $\sigma \equiv (\sigma_{1} + \sigma_{2})/2$. In terms of the above quantities, Eqs. (\ref{3}) and (\ref{11}) become
\begin{equation}
\label{14}
a_{i} \theta_{i} + \frac{b_{i}}{2N_{i} d}\, \left( \theta_{i}-\theta_{P} \right)- c_{i} \theta_{i}^{1/2}=0,
\end{equation}
\begin{equation}
\label{15}
\rho \left( \theta_{1} -\theta_{P} \right) + \left( \theta_{2} - \theta_{P} \right) =0,
\end{equation}
$i=1,2$. In the above equations,
\begin{equation}
\label{16}
a_{i} \equiv \frac{\pi^{d/2} \zeta^{*}(\alpha_{i})}{\sqrt{2} (d+2) \Gamma (d/2)}\, \left( \frac{\sigma_{i}}{\sigma} \right)^{d-1} \left( \frac{m}{m_{i}} \right)^{1/2}\, ,
\end{equation}
\begin{equation}
\label{17}
b_{i} \equiv \frac{(m m_{i})^{1/2}S_{P}}{M \sigma^{d-1}},
\end{equation}
\begin{equation}
c_{i} \equiv \frac{ \sqrt{\pi} S_{Wi}}{4 \sqrt{2}d \sigma^{d-1}N_{i}}\, ,
\end{equation}
and
\begin{equation}
\label{18}
\rho \equiv \left( \frac{m_{1}n_{1}}{m_{2}n_{2}} \right)^{1/2}\, .
\end{equation}
Equations (\ref{14}) and (\ref{15}) together with Eq.\ (\ref{2}) rewritten as
\begin{equation}
\label{19}
\rho^{2} \frac{m_{2}}{m_{1}}\, \theta_{1}=\theta_{2},
\end{equation}
define a closed set of equations for the quantities $x_{P}$, $\theta_{1}$, $\theta_{2}$, and $\theta_{P}$. It is worth to remind that $c_{i}$ depends on $x_{P}$ through $S_{Wi}$. More specifically, it is
\begin{equation}
\label{20}
S_{W1}= S_{P}+\gamma x_{P}, \quad S_{W2}= S_{P}+\gamma (L_{x}-x_{P}),
\end{equation}
with $\gamma$ being some geometrical factor with dimensions of $\mbox{(length)}^{d-2}$. For instance, if the system is a circular cylinder $\gamma  =2 (\pi S_{P})^{1/2}$, while for a two-dimensional rectangular system it is $\gamma=2$.

Since the problem has many parameters, let us particularize for the case in which all the particles are identical and there is the same number of them in each compartment, i.e. $N_{1}=N_{2}=N$, $\alpha_{1}=\alpha_{2}=\alpha$, $\sigma_{1}= \sigma_{2} =\sigma$, $m_{1}=m_{2}=m$. Then, after some algebra, from Eqs.\ (\ref{14}), (\ref{15}), and (\ref{20}), it is possible to derive a closed equation for $\rho$,
\begin{equation}
\label{21}
\left(1- \rho \right) \left(\rho^{2}+\lambda \rho+1 \right)=0,
\end{equation}
\begin{equation}
\label{22}
\lambda \equiv \frac{2N \gamma L_{x} ad}{2 S_{P} \left( Nad+b \right) +b \gamma L_{x}},
\end{equation}
where $a=a_{1}=a_{2}$ and $b=b_{1}=b_{2}$. Note that now it is
\begin{equation}
\label{23}
\rho = \left( \frac{n_{1}}{n_{2}} \right)^{1/2} = \left( \frac{L_{x}}{x_{P}}-1 \right)^{1/2}\, .
\end{equation}
Once the density parameter $\rho$ is known, the value of the temperature $\theta_{1}$ can be obtained by combination of Eq. (\ref{14}) for $i=1$ and $i=2$ and using Eq. (\ref{15}). The result is
\begin{equation}
\label{24}
\theta_{1}= \frac{\pi}{32} \left[ \frac{2S_{P} + \gamma L_{x}}{a N \sigma^{d-1} d} \right]^{2} \frac{1}{(1+\rho)^{2}}.
\end{equation}
Moreover, the temperature parameters are related by
\begin{equation}
\label{25}
\theta_{P} = \rho \theta_{1} = \left( \theta_{1} \theta_{2} \right)^{1/2}.
\end{equation}
Equation (\ref{21}) has the expected symmetric solution $\rho=1$, i.e. $n_{1}=n_{2}$, $x_{P}=L_{x}/2$. Also, it follows from Eqs. (\ref{15}) and (\ref{19}) that $\theta_{1}=\theta_{2}= \theta_{P}$, so that the gases in both compartments and the piston all have the same temperature parameter. This state exhibits the symmetry of the material parameters at both sides of the piston. But Eq.\ (\ref{21}) has also  another solution corresponding to an asymmetric state, in which the steady position of the piston is not in the middle of the system but at a position defined by
\begin{equation}
\label{26}
\rho_{\pm} = \frac{1}{2} \left( \lambda \pm \sqrt{\lambda^{2}-4} \right).
\end{equation}
The two solutions $\rho_{+}$ and $\rho_{-}= \rho_{+}^{-1}$ are obtained one from the other by interchanging the two compartments. The existence of this asymmetric state requires that the parameters of the system verify the condition $\lambda >2$. It follows from Eq. (\ref{19}) that the steady temperatures of the granular gases in the two compartments are different in the asymmetric state.

To verify the accuracy of the above theoretical predictions, molecular dynamics (MD) simulations have been performed
for a system of $2N=200$ equal inelastic hard disks ($d=2$) enclosed in a rectangular container of size $L_{x} \times L_{y}$. The latter corresponds to the section $S_{P}$ used above. All the simulations to be reported in the following started from a symmetric situation with the piston at rest in the middle, $x_{P}(0)=L_{x}/2$, and equal number of particles in both compartments. The initial velocity distributions were Gaussian and with the same temperature. Finally, the values of $L_{x}$ and $L_{y}$ were always chosen such that the initial number density was quite low.

\begin{figure}
\onefigure[scale=0.7]{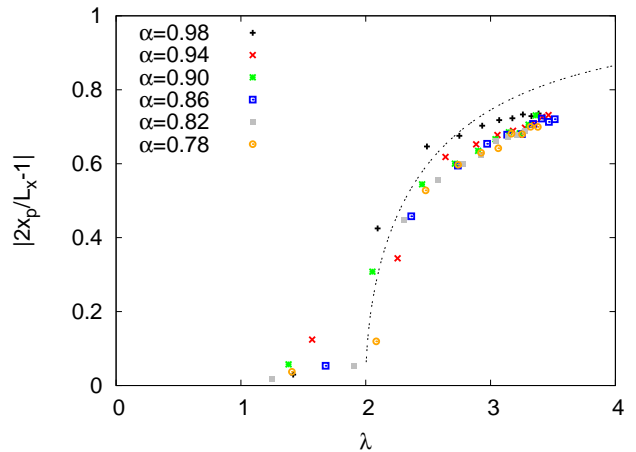}
\caption{Bifurcation diagram showing the asymmetry of the steady position of the piston as a function of the dimensionless control parameter $\lambda$ defined in Eq.\ (\protect{\ref{22}}). The symbols are from molecular dynamics simulations and the solid line is the theoretical prediction given by Eq. (\protect{\ref26}) with $\rho=\rho_{+}$. The values of the parameters are $L_{x}=2L_{y}= 133 \sigma$ and $2N=200$, i.e. $n\sigma^{2}=0.02$. Different values of $\alpha$ have been used as indicated in the insert.
In addition, also different values of the ratio $M/m$ have been employed to change the value of $\lambda$ for each value of $\alpha$.}
\label{fig2}
\end{figure}

\begin{figure}
\onefigure[scale=0.7]{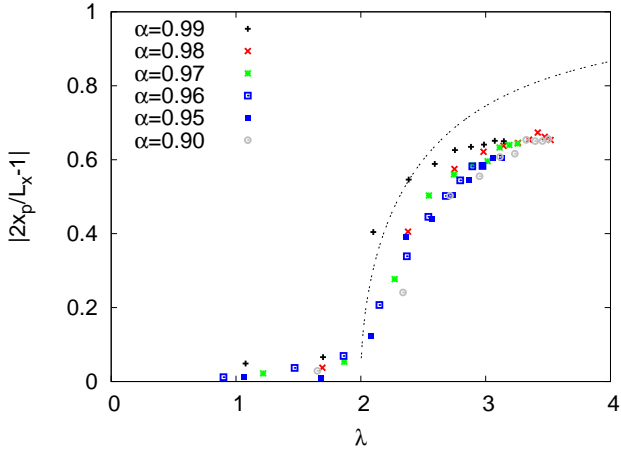}
\caption{The same as in Fig.\ \protect{\ref{fig2}}, but for a system with $L_{x}=2L_{y}= 100 \sigma$, so that $ n \sigma^{2} =0.04$.}
\label{fig3}
\end{figure}

As predicted by the theory developed above, for small values of $\lambda$ defined in Eq. (\ref{22}), the piston remains oscillating around  the middle of the system. Nevertheless, when the value of $\lambda$ increases above $2$, the average position of the piston clearly moves away from the initial position until reaching a different steady average. In this way, the symmetry of the system is broken. In Fig.\ \ref{fig2}, the steady average position of the piston is plotted as a function of the dimensionless control parameter $\lambda$ for a system with $L_{x}=2L_{y}= 133 \sigma$, so that the initial homogeneous  density in both compartments is $n=0.02 \sigma^{-2}$.  More precisely, the plotted quantity is
\begin{equation}
\label{27}
\epsilon \equiv \frac{| 2x_{P}-L_{x}|}{L_{x}} = \frac{|1-\rho^{2}|}{\rho^{2}+1}.
\end{equation}
Different values of the coefficient of normal restitution $\alpha$ have been considered. Moreover, for each value of $\alpha$, the values of $\lambda$ has been modified by changing the ratio $M/m$ between the mass of the piston and the mass of the particles. A reasonably good agreement is found between theory and simulations, the former being given by the last equality  in Eq.\ (\ref{26}) with $\rho=\rho_{+}$, as given by Eq.\ (\ref{25}). It must be taken into account that when the system is above but close to the bifurcation point, the position of the piston does not oscillate around a unique steady position but also moves back and forth between the two possible ones, at both sides of the middle of the system. It is not an easy task to separate both motions as required to compare with the theory developed here, which does not consider fluctuations between the two steady positions. Similar results have been obtained for other values of the average number density, as illustrated in Fig. \ref{fig3} for $n=0.04 \sigma^{-2}$. As expected, the quantitative accuracy of the theoretical prediction decreases as the density increases, since the model has been formulated in the low density limit. Let us also notice that for the geometry  used in our simulations, when $L_{x}  \leq L_{y}$, it is $\lambda <2$ independently from the values of the other parameters of the system and, therefore, the theory predicts that there is no spontaneous symmetry breaking. This feature has been consistently confirmed by the simulation results.

\begin{figure}
\onefigure[scale=0.7]{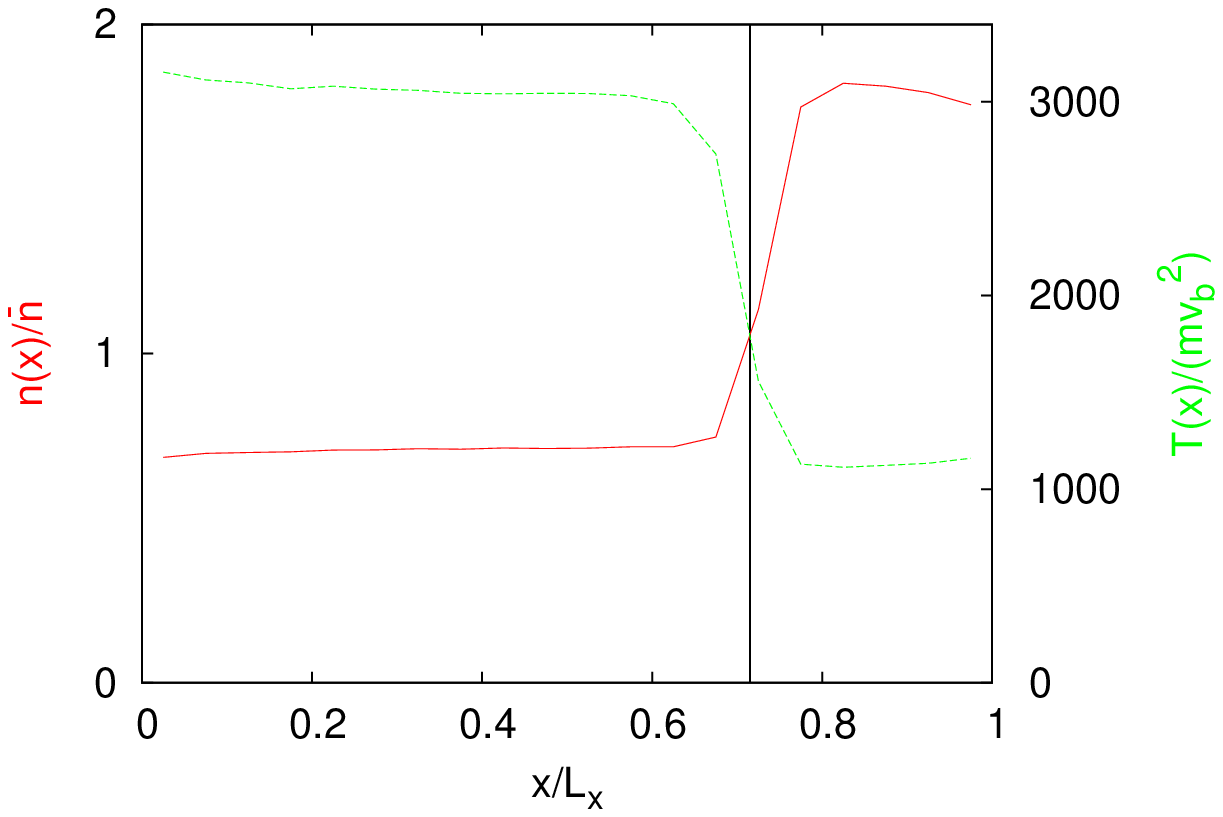}
\caption{Dimensionless steady number density $n(x)/n$, $n= N/L_{x}L_{y}$, (solid line and left axis) and temperature $\theta(x) \equiv T(x)/ mv_{b}^{2}$  (dashed line and left axis) profiles along the $x$-direction for a system with $L_{x}=2L_{y}= 100 \sigma$, $\alpha=0.99$, and $M=150 m$. The vertical line indicates the theoretical prediction for the average position of the piston.   }
\label{fig4}
\end{figure}

It is interesting to verify at what extent the hypothesis on which the theory formulated above is based are verified and, in particular, the homogeneity of the subsystems. To measure the temperature and number density fields in the simulations, the system has been divided into cells parallel to the piston of width $L_{x}/20$. The fields have been measured once the system has reached the steady state, and the reported profiles have been averaged over time and also over trajectories. In Fig. \ref{fig4} the steady average density and temperature profiles, $n(x)$ and $T(x)$, are shown for a system with  $L_{x}=2L_{y}=100 \sigma$, and $\alpha=0.99$, and $M=150m$. This gives $\lambda \simeq 2.1$ and the prediction for the steady position of the piston is $x_{P} \simeq 0.65 L_{x}$. It is observed that outside from  a boundary layer at both sides  of the piston, the hydrodynamic fields can be considered as uniform inside each compartment, although having quite different values in each one. The boundary layers are, at least partially, due to the oscillations of the piston.

\begin{figure}
\onefigure[scale=0.7]{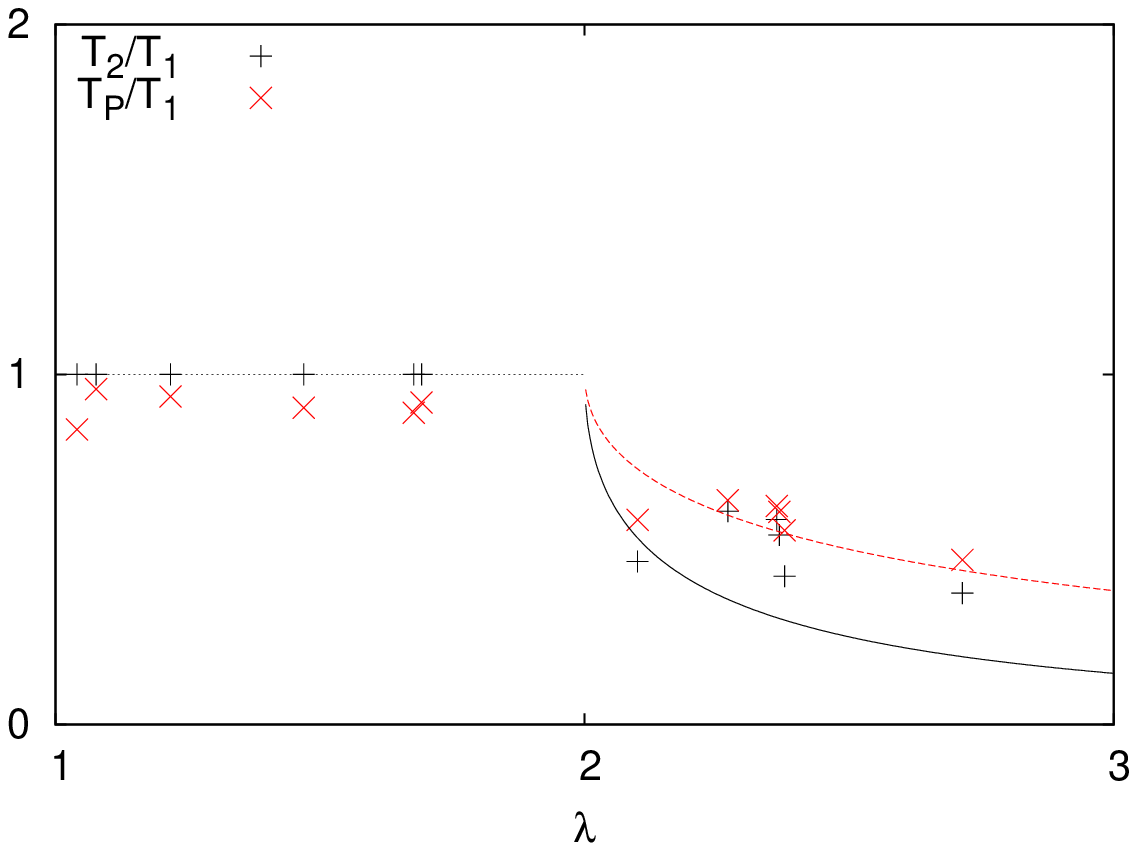}
\caption{Ratio of the granular temperatures of the gases in both compartments $T_{2}/T_{1}$ (solid line and plus symbols) and between the temperature parameter of the piston and that of compartment 1, $T_{P}/T_{1}$ (dashed line and crosses) as a function of the control parameter $\lambda$, defined in Eq.\ (\protect{\ref{22}}).  The lines are the theoretical predictions derived here while the symbols are molecular dynamics simulation results. The size of the system is $L_{x}=2 L_{y} =100 \sigma$.}
\label{fig5}
\end{figure}

Another quantitative prediction of the model are the values of the temperature ratios $T_{2}/T_{1}= \theta_{2}/\theta_{1}$ and
$T_{P}/T_{1} = \theta_{P} / \theta_{1}$. The theoretical predictions are given by Eqs.\ (\ref{20}) and (\ref{26}). They are compared with the simulation results for a system with $L_{x}=L_{y} = 100 \sigma$ in Fig.\ \ref{fig5}. When the system is in an asymmetric state, the temperature of each of the compartments has been measured by omitting the boundary layers next to the piston. Again, given the approximations involved in the formulation of the model, the accuracy can be considered as satisfactory.

In summary, we have found  a spontaneous symmetry breaking in two vibrated granular gases separated by an adiabatic  piston.
Moreover, the energy balance equations (\ref{3}) and (\ref{11}), based on simple kinetic theory arguments, capture surprisingly well the qualitative and quantitative features observed in the molecular dynamics simulations.

Of course, a relevant question is to know which is, if any, the macroscopic criterion leading to equilibration of the granular gases in both compartments. In this sense, a first conclusion of the analysis here is that there is a macroscopic description of
the steady states of the system in terms of the densities of both compartments and {\em one} temperature parameter. From them, the other two temperature parameters and the pressure can be determined, if needed. Also, it is possible to compare  the energy dissipation in the symmetric and asymmetric steady states. The power $P_{D}$ dissipated in collisions in a steady state is
\begin{eqnarray}
\label{28}
P_{D} & = & \frac{N d}{2} \left( \zeta_{1} T_{1} + \zeta_{2} T_{2} \right) \propto n_{1}T_{1}^{3/2} + n_{2} T_{2}^{3/2}
\\ \nonumber
& \propto &
\frac{\theta_{1}^{3/2}}{x_{P}}
+ \frac{\theta_{2}^{3/2}}{L_{x}-x_{P}}
\propto \frac{\rho^{2}+1}{(1+\rho)^{2}}\ .
\end{eqnarray}

This function has a minimum for $\rho=1$, i.e. the symmetric state. This shows that the system chooses between the allowed states
the one  having a maximum  energy dissipation in the granular gases. This information could be relevant to fromulate some extremal principle governing the macroscopic evolution of granular fluids.

\acknowledgments

This research was partially supported by the Ministerio de
Educaci\'{o}n y Ciencia (Spain) through Grant No. FIS2008-01339 (partially financed
by FEDER funds).

\end{document}